\documentclass[conference]{IEEEtran}
\IEEEoverridecommandlockouts
\usepackage{cite}
\usepackage{amsmath,amssymb,amsfonts}
\usepackage{algorithmic}
\usepackage{graphicx}
\usepackage{textcomp}
\usepackage{xcolor}
\def\BibTeX{{\rm B\kern-.05em{\sc i\kern-.025em b}\kern-.08em
    T\kern-.1667em\lower.7ex\hbox{E}\kern-.125emX}}
\begin{document}
\title{A new weighted ensemble model for phishing detection based on feature selection\\
{\footnotesize }
\thanks
}
\author{\IEEEauthorblockN{1\textsuperscript{st}Farnoosh Shirani Bidabadi}
\IEEEauthorblockA{\textit{dept. Computer Science} \\
\textit{Texas A\&M University}\\
College Station,USA \\
farnoosh.shirani@tamu.edu}
\and
\IEEEauthorblockN{2\textsuperscript{nd}Shuaifang Wang}
\IEEEauthorblockA{\textit{dept. Computer Science} \\
\textit{Texas A\&M University}\\
College Station,USA \\
wangshuaifang@tamu.edu}
}
\maketitle

\begin{abstract}
A phishing attack is a sort of cyber assault in which the attacker sends fake communications to entice a human victim to provide personal information or credentials. Phishing website identification can assist visitors in avoiding becoming victims of these assaults. The phishing problem is increasing day by day, and there is no single solution that can properly mitigate all vulnerabilities, thus many techniques are used. In this paper, We have proposed an ensemble model that combines multiple base models with a voting technique based on the weights. Moreover, we applied feature selection methods and standardization on the dataset effectively and compared the result before and after applying any feature selection.
\end{abstract}

\section{Introduction}
Phishing is a type of online fraud attempts to trick people into providing sensitive information such as credit card numbers, usernames, and passwords. It typically is done by sending users emails that contain a hyperlink to a fake website. According to [4], more than 255 millions phishing attacks are reported in 2022, which is a 61\% rate of increase compared to 2021. Therefore, phishing attacks are evolving at a quicker pace than many of us seem to realize and it is necessary to develop a more efficient anti-phishing tool. This paper proposed a method for the identification of phishing websites through an ensemble of 5 different machine learning classifiers Random Forest(RF), K-Nearest Neighbor(KNN), Support Vector Machine(SVM), Logistic Regression(LR), and NaiveBayes(NB) with a top-rank feature selection technique. In deciding on whether to keep a certain feature, we applied different feature selection techniques such as correlation and information gain to sort the feature by importance and came up with an appropriate n value to reduce feature sets. From there we trained every base model with reduced feature sets and built an ensemble model using the weighted voting technique. Results are measured by accuracy as well as precision, recall, and f1-score that were obtained based on the confusion matrix. The proposed model has shown promising results in terms of prediction accuracy. Overall, the purpose of this study is to propose a generalized new approach that is built on machine-learning algorithms in today's analytics platforms. 

\section{Related Work}
Lots of previous research has been put into developing an anti-phishing technique. Our literature review indicated that feature selection and ensemble models are widely used by many authors in order to develop an efficient phishing detection tool. There are several versions of research that focus on feature selection and ensemble machine learning base model. Paper[3] presented a machine learning based technique comprising of feature selection module to increase the effectiveness of model performance. Two well known classifiers Naive Bayes(NB) and Support Vector Machine(SVM) were applied to train a 15-dimensional feature set. The model achieved an accuracy of 99.96\%. According to paper[1], the authors proposed an idea of selecting top-n feature subset to improve the performance of each base model. Noticing that in order to reach the best performance, the number of features included in each algorithm varies in every scenario, leading to a conclusion that the proposed model could not lessen the number of features in a significant way. Almseidin et al. [8] published research in which they improved the effectiveness of their system by employing several machine learning algorithms and feature selection approaches. The studies were carried out using a phishing dataset of 48 attributes, which included 5000 legitimate and 5000 phishing webpages. They came to the conclusion that an RF algorithm with only 20 characteristics provides the greatest accuracy. In paper[2], a ensemble model was built on individual decision trees with proposed RRFST. Although the model produced a remarkable accuracy of 99.27\%, having only one classifier in an ensemble model may not be considered as a general approach. In other words, the proposed method works under the assumption that users have the prior knowledge of which model outperforms others. Thus, this proposal stands out as a better approach as our ensemble model is based on each weighted base model with reduced feature subset. We believe that our method is more generalized and result-oriented than prior research. 

\section{Methodology}
This section describes the method used for distinguishing legitimate webpages from phishing ones. five machine-learning models are applied to the dataset, then an ensemble model is built that combines multiple base models with the weighted voting technique.
Following that, a feature selection approach is used to choose top-rank features. These selected features are used in the classification models and the performance of the models is compared before and after applying feature selection.

\subsection{Dataset}\label{AA}
The dataset of this proposal comes from Kaggle, which includes 11430 URLs with 87 extracted features. Features are from three different classes: 56 are extracted from the structure and syntax of URLs, 24 are extracted from the content of their correspondent pages, and 7 are extracted by querying external services. The dataset is balanced, it contains exactly 50\% phishing and 50\% legitimate URLs. In terms of data preprocessing, we noticed that one of the columns status is of type string so we decided to drop that column and make a new column of integer type, with 1 representing phishing and 0 representing legitimate. By doing this we can label the data in a more intuitive way which helps us in further training and testing. 

\subsection{core algorithms}

In this paper, we have applied five machine learning classifiers, namely Random Forest (RF), K-Nearest Neighbor(KNN), Support Vector Machine(SVM), Naive Bays(NB), and Logistic Regression(LR). Afterward, these five base models are weighted based on their performance and an ensemble model is built with voting that aggregates the prediction of each base model.

\subsection{Feature Selection and Standardization}
A feature selection approach is used to choose the top-ranked features, which aids in the removal of irrelevant ones. Building a classification model with fewer but more important features aids in enhancing learning speed and performance. In this work, the information gain (IG) [9] method is used for selecting the top features. Information gain indicates how much information a given feature gives us about the final outcome. It can be calculated by subtracting the entropy of a specific attribute within the data set from the entropy of the entire data set.
\begin{equation}
Entropy=-\sum_{i}P{_i}log{_2}P{_i}
\end{equation}
where $P{_i}$ is the probability of the class.

Figure 1 shows the top 20 attributes after applying IG and the ranking algorithm. These selected attributes are used for classification.
\begin{figure}[htbp]
\centerline{\includegraphics[width=0.5\textwidth]{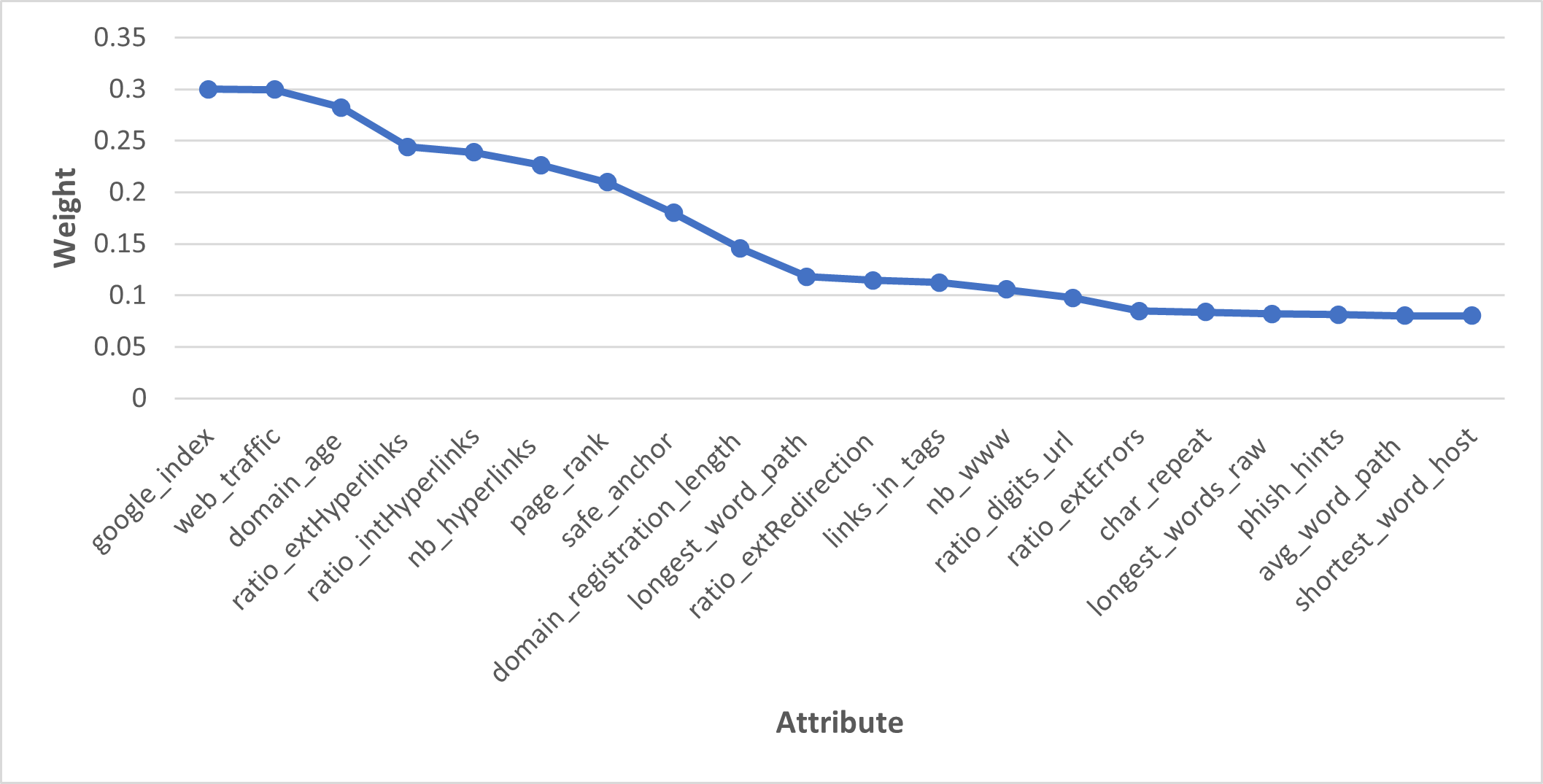}}
\caption{Top 20 attribute ranking.}
\label{fig}
\end{figure}
Variables assessed at various scales may not contribute equally to the model fitting and learning function and may result in bias. As a result, to address this possible issue, features have been standardized. It removes the mean and scales each feature to unit variance.
\subsection{ensemble model}
Once we determined the top 20 important features, an ensemble model was built to combine all 5 base models to boost the overall performance. As shown from figure 2, specifically we implemented the weighted strategies that calculate the weighted average probability from individual models using train AUC as weights. The weights were later served as an input to the voting classifier that combines probabilities of each prediction and picks the prediction with the highest total probability. 
\begin{figure}[htbp]
\centerline{\includegraphics[width=0.5\textwidth]{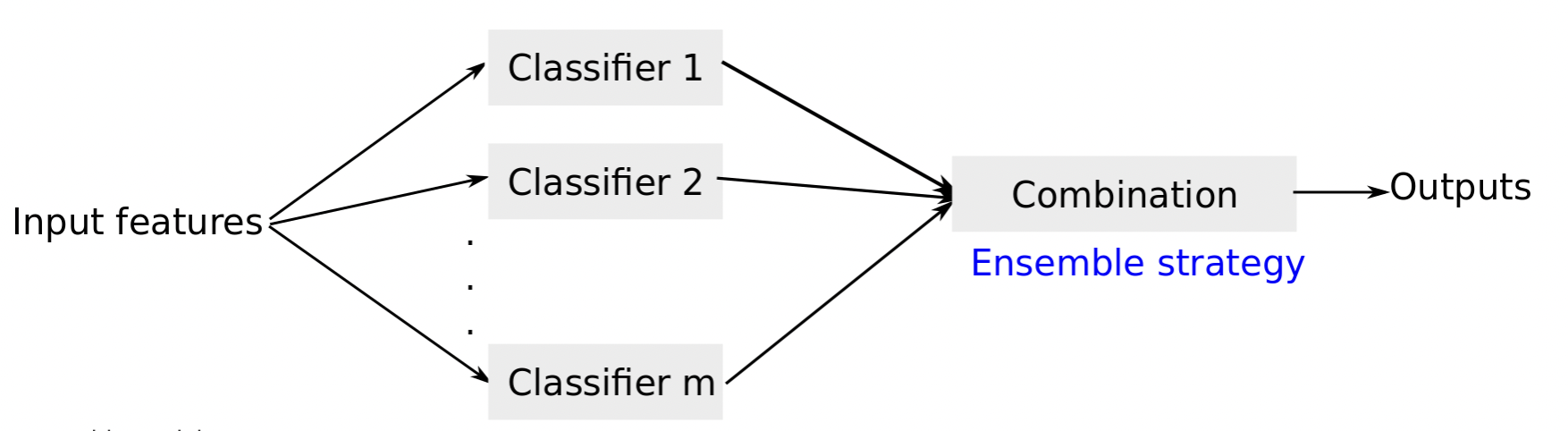}}
\caption{Ensemble model}
\label{fig}
\end{figure}

\section{Evaluation Metrics}
In order to evaluate machine learning algorithms, we have used various evaluation parameters. These are precision, recall, F1-score, accuracy. These are computed by using the fields of the confusion matrix as described in Table I.
\begin{table}[h!]
\caption{Confusion Matrix feilds}
\centering
\begin{tabular}{|l|l|}
\hline
\textbf{Name}       & \textbf{Description}                             \\ \hline
True Positive(TP)   & No. of phishing webpages predicted correctly     \\ \hline
False negative (FN) & No. of phishing webpages predicted incorrectly   \\ \hline
False positive (FP) & No. of legitimate webpages predicted incorrectly \\ \hline
True negative (TN)  & No. of legitimate webpages predicted correctly   \\ \hline
\end{tabular}
\end{table}
\begin{itemize}
\item \textbf{Precision:} This measurement represents a degree of exactness.
\begin{equation}
Precision=\frac{TP}{TP+FP}
\end{equation}
\item \textbf{Recall:} It is defined as the rate of phishing webpages that are correctly identified.
\begin{equation}
Recall=\frac{TP}{TP+FN}
\end{equation}
\item \textbf{F1-Score :} It is the harmonic mean of precision and recall.
\begin{equation}
F1-Score =\frac{2*Precision*Recall}{Precision+Recall}
\end{equation}
\item \textbf{Accuracy(\%):}  It is the percentage of correctly recognized phishing and legitimate webpages.
\begin{equation}
Accuracy=\frac{TP+TN}{TP+FN+TN+FP}*100 
\end{equation}\\
\end{itemize}
\section{Experiments and Results}
This section includes the experimental results, analyses, and visualizations.

We first applied 5 base models on all extracted features. Table below shows the output metrics for each model. RF classifier has demonstrated the best result with an accuracy of 96.7\%, while SVM model has the worst performance of 44.6\% accuracy. 
\begin{table}[h!]
\caption{Classification metrics before feature selection and standardization}
\centering
 \begin{tabular}{||c c c c c||} 
 \hline
 Classifier & Accuracy(\%)  & Precision & Recall & F1-Score \\ [0.5ex] 
 \hline\hline
 RF & 96.7 & 0.964 & 0.968 & 0.966 \\
 SVM & 44.6 & 0.399 & 0.225 & 0.288 \\
 KNN & 82.4 & 0.833 & 0.808 & 0.820 \\
 LR & 77.5 & 0.752 & 0.819 & 0.784 \\
 NB & 87.4 & 0.863 & 0.887 & 0.875 \\ [0.5ex] 
 \hline
 \end{tabular}
\end{table}

Afterwards, feature selection and standardization has applied on the same models as shown on Figure 3, the accuracy of SVM improved significantly.
\begin{table}[h!]
\caption{Classification metrics after feature selection and standardization}
\centering
 \begin{tabular}{||c c c c c||} 
 \hline
 Classifier & Accuracy(\%) & Precision & Recall & F1-Score \\ [0.5ex] 
 \hline\hline
 RF & 96.11 & 0.960	& 0.953 &	0.956\\
 SVM & 91.64 & 0.899 & 0.934 &	0.916 \\
 KNN & 94.93 & 0.943 & 0.944 &	0.944\\
 LR & 93.04 & 0.930	& 0.935 &	0.933\\
 NB & 91.56 & 0.894 & 0.918	 & 0.906\\ [0.5ex]
 \hline
 \end{tabular}
\end{table}
\begin{figure}[htbp]
\centerline{\includegraphics[width=0.5\textwidth]{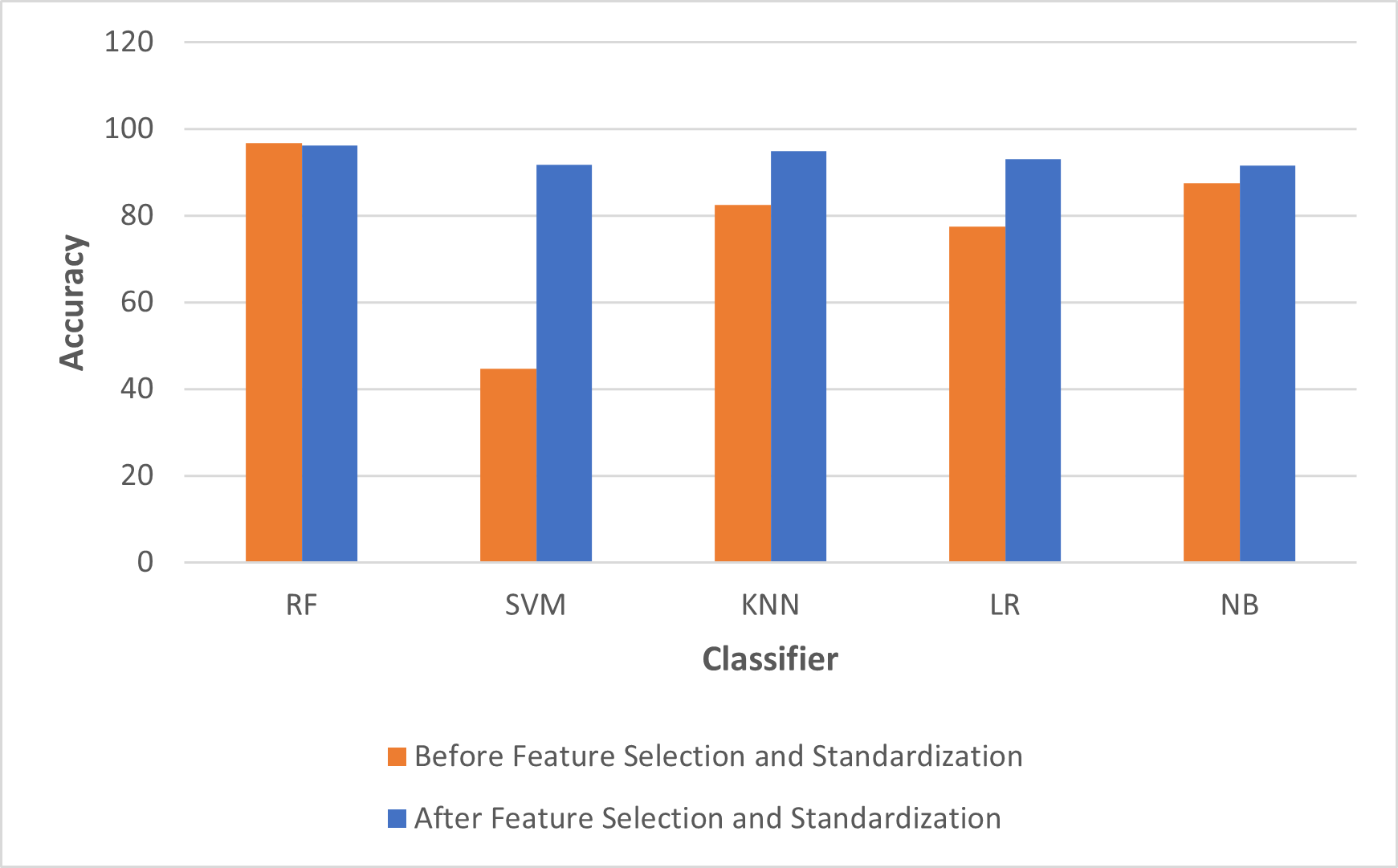}}
\caption{Comparison accuracy of classifiers before and after feature selection and standardization}
\label{fig}
\end{figure}

Once we implemented feature selection, we built an ensemble model with base models using weighted voting technique. The performance of ensembling model is shown in Figure 4. \\
We also created other ensemble model such as bagging regression, however given the fact that random forest algorithm is considered an extension of the bagging method, we don't think it is an optimal method to implement. Weight was passed in as a parameter to the voting classifier and it stores the accuracy of individual base model on selected features.

\begin{figure}[htbp]
\centerline{\includegraphics[width=0.5\textwidth]{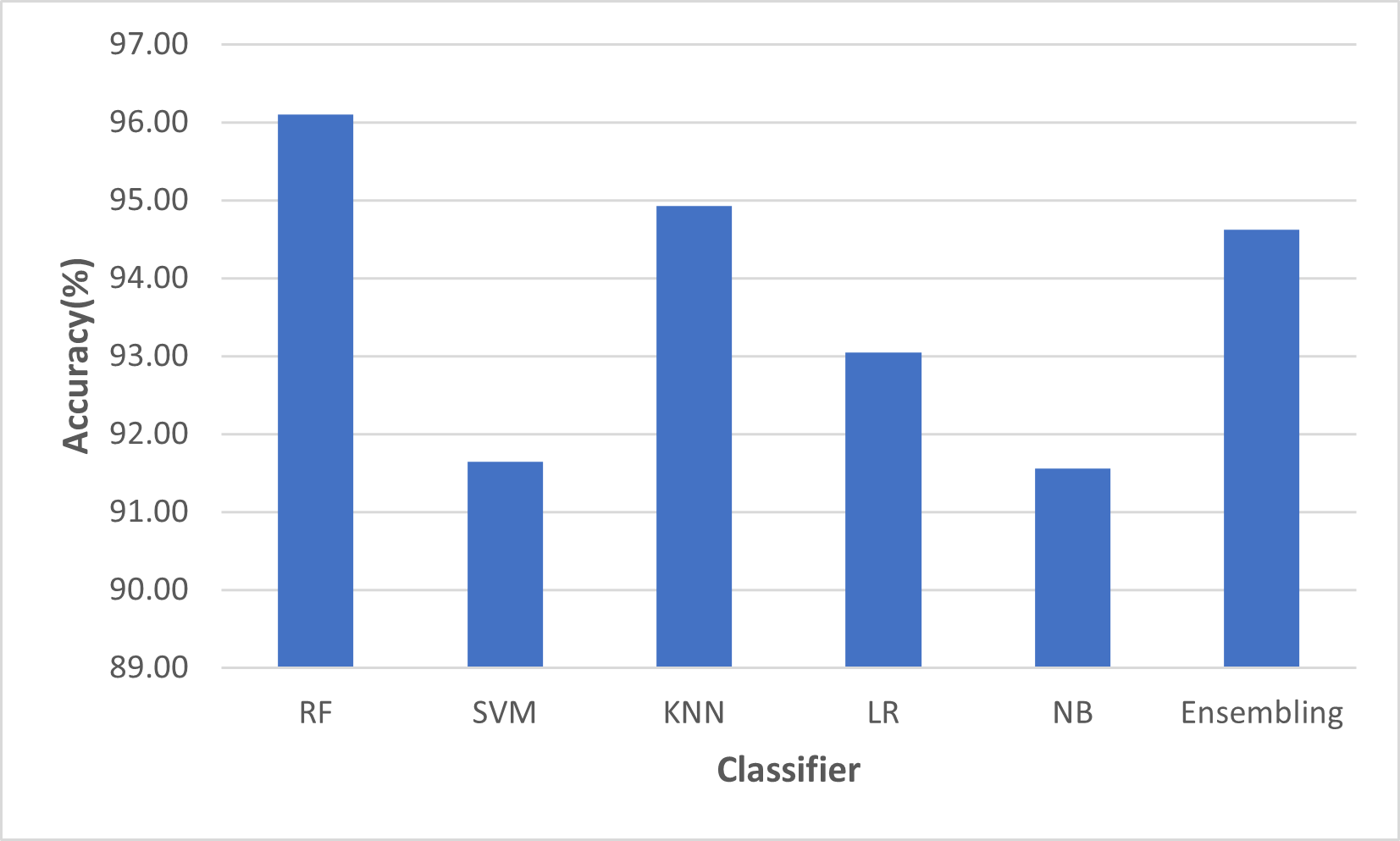}}
\caption{Performance of base models and ensemble model after feature selections}
\label{fig}
\end{figure}

\section{Conclusion and future Expectations}
This paper described an effective approach for detecting phishing websites. For classification, it employs a feature selection strategy in addition to machine learning methods. The result of all 5 classifiers is compared before and after feature selection and standardization. Moreover, the performance of the ensemble model is studied in this paper. The experimental findings clearly show that using a feature selection strategy to choose a relevant collection of features increases classification model performance. 
Moreover, we can notice that ensemble models do not always improve accuracy. It certainly helps in reducing the variance of multiple models by aggregating all classifiers together and thereby improving the average performance. However, there are cases where an individual model can outperform a group of other models. In this case, random forest performs better than other models and the ensemble model seems to be beneficial only when all classifiers perform at similar levels. Therefore, by knowing the information that one algorithm always outperforms others in advance, it is reasonable to make an assumption that the specific algorithm is adequate to make predictions for our dataset.

\newpage
{\small
\bibliographystyle{ieee_fullname}
\bibliography{egbib}
}

\vspace{12pt}
\end{document}